**Author for corrspondance**
Dr. P. Jonnard
Laboratoire de Chimie Physique-Matière et Rayonnement, 11 rue Pierre et Marie Curie
F-75231 Paris Cedex 05, FRANCE
Tel: 33 1 44 27 63 03 fax : 33 1 44 27 62 26 e-mail : philippe.jonnard@upmc.fr

# High-resolution x-ray analysis with multilayer gratings

***P. Jonnard, K. Le Guen, J.-M. André***

Laboratoire de Chimie Physique - Matière et Rayonnement, Univ UPMC Paris 06, CNRS - UMR 7614, 11 rue Pierre et Marie Curie, F-75231 Paris Cedex 05, FRANCE

## Abstract

Periodic multilayers are nowadays widely used to perform x-ray analysis in the soft x-ray range (photon energy lower than 1 keV). However, they do not permit to obtain high-resolution spectra like natural or synthetic crystals. Thus, multilayers cannot resolve interferences between close x-ray lines. It has been shown and demonstrated experimentally that patterning a grating profile within a multilayer structure leads to a diffractive optics with improved resolving power. We illustrate the use of a $Mo/B_4C$ multilayer grating in the Fe L and C K spectral ranges, around 700 eV and 280 eV respectively. First, in the Fe L range, the improved spectral resolution enables us to distinguish the Fe Lα and Lβ emissions (separated by 13 eV). In addition, using a sample made of a mix of LiF and an iron ore, we show that it is possible to easily resolve the F K and Fe L emissions. These examples demonstrate that an improved x-ray analysis can be obtained with multilayer gratings when there is the need to study samples having elements giving rise to close emission lines. Second, in the C K range, by comparing C Kα spectra from $B_4C$ and cellulose, we show that the shape of the emission band is sensitive to the chemical state of the carbon atom.

# Introduction

Periodic multilayers, consisting in a stack of bilayers with alternating light and heavy materials of nanometric size, are nowadays widely used to perform wavelength dispersive spectrometry in the soft x-ray range [1-3]. Indeed, their period can be adjusted for the wavelengths larger than 3 nm. However, multilayers suffer from moderate spectral resolution due to their large diffraction patterns. With multilayers as analyser elements, the quantitative analysis is often difficult or even impossible, in particular in case of two close emission lines. It has been proposed [4] and experimentally demonstrated [5,6] that it is possible to improve the resolution by etching the multilayers according to the profile of a lamellar grating. Thus, depending on the spectral range and on the grating parameters (period and size of the grooves) an improvement of the spectral resolution by factor 2-3 can be obtained to the detriment of a small loss of reflectivity.

In this paper, we present the use of a Mo/$B_4C$ multilayer grating in the C K and Fe L spectral ranges. For each range, we study two different samples and compare the results with those obtained with a Mo/$B_4C$ multilayer, in order to discriminate the performances of both diffracting structures in terms of spectral resolution but also peak and background intensities.

# Experiment

The samples are successively analysed with a periodic multilayer and a multilayer grating etched in the same multilayer. The multilayer is made of 150 Mo/$B_4C$ bilayers deposited onto a silicon substrate. The thickness of both Mo and $B_4C$ layer is 3 nm giving a multilayer period of 6 nm. The period of the grating is 1.0 µm and the width of the grooves is 0.7 µm; thus the width of the multilayer bar is 0.3 µm. The preparation and characterization of the multilayer grating are described in Refs. [5,7-9]. These structures have been initially designed to give optimal performances in the B K spectral range (around 6.7 nm or 180 eV). For spectroscopic purposes, the multilayer gratings are used at the first, second, …, diffraction order of the multilayer and at the zeroth order of the grating.

The multilayer and multilayer grating are placed in a home-made spectro-goniometer [10], used in the present case as a plane x-ray spectrometer. The different samples are successively placed inside a windowless x-ray tube and used as the anode. The energy of the exciting electrons is 2 keV with a current of 10-40 mA depending on the samples. During the experiments, the pressure ranges within $10^{-5}$-$10^{-6}$ mbar inside the apparatus.

Two samples are studied in the Fe L spectral range: an iron metal plate and a powder consisting in a mix of LiF and an iron ore powders. Two samples are studied in the C K spectral range: a $B_4C$ thin film and a cellulose pellet. It is observed after the electron excitation that the insulating samples (mixed powder and cellulose) have been degraded.

## X-ray analysis in the Fe L range

We first present the x-ray analysis of the iron metal target. Its Fe L spectrum obtained with the multilayer and the multilayer grating is presented in Figure 1. The spectra are obtained with the same excitation conditions. A large intensity decrease is noted when using the multilayer grating. This is because this one is not optimized to work in this spectral range. The maxima are not at the same position because the refraction index is not the same for the two diffracting structures. In fact, the refraction indexes are different between the multilayer and the grating because there are many empty spaces in the multilayer grating. This leads to different optical paths with the structures and then to different effective periods.

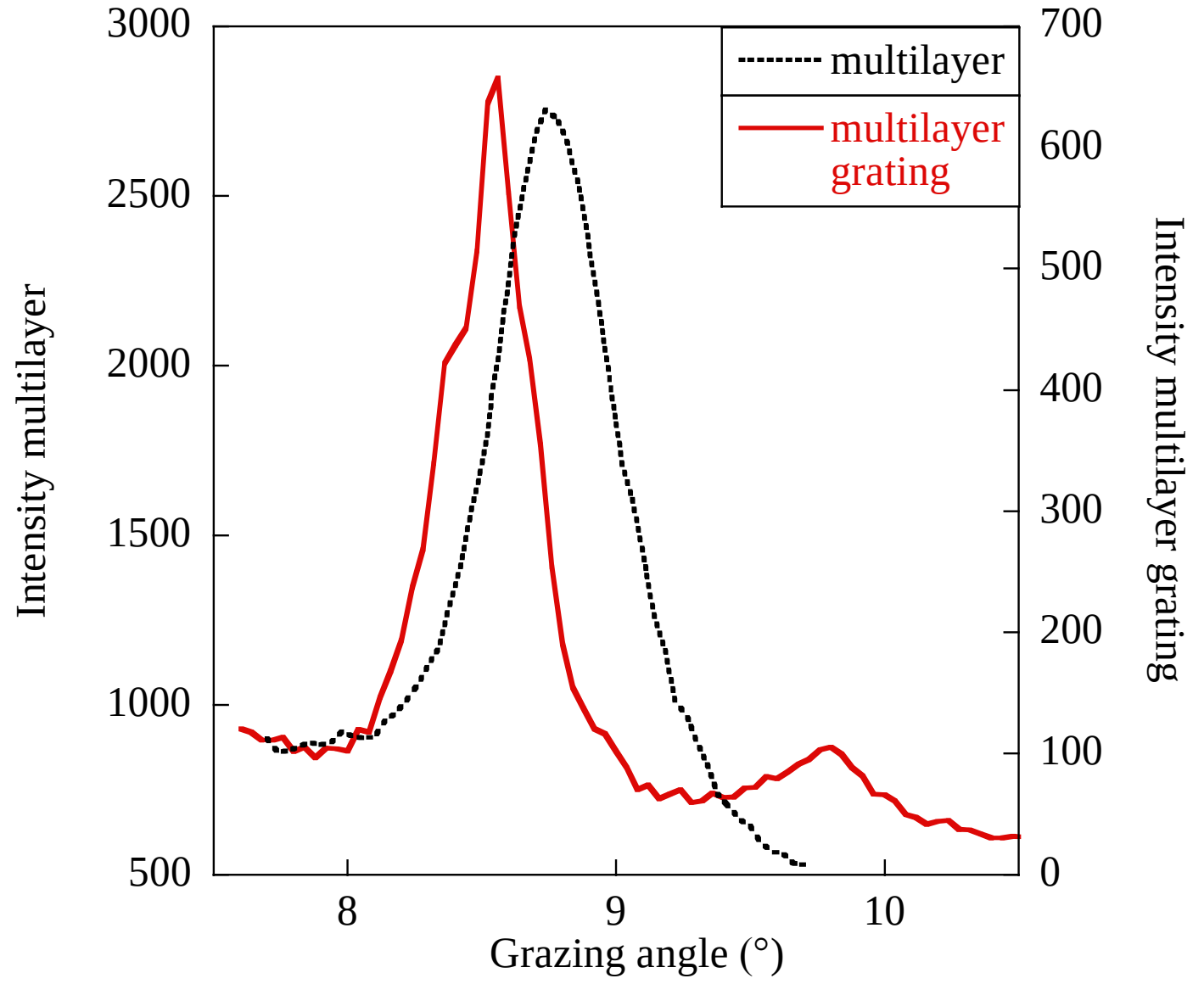

**Figure 1**: Fe L spectral range of an iron target analysed with a Mo/$B_4C$ multilayer and the Mo/$B_4C$ multilayer grating.

It is clearly observed in Figure 1 that the full width at half maximum (FWHM) is narrower with the multilayer grating (0.39°) than with the multilayer (0.53°). This enables us to distinguish the Fe Lβ (3d – $2p_{1/2}$ transition) emission as a shoulder around a Bragg angle of 8.4°, the maximum of the Fe Lα emission (3d – $2p_{3/2}$ transition) being located at 8.5°. This angular difference corresponds to an energy difference of about 13 eV in agreement with the $2p_{1/2} – 2p_{3/2}$ spin orbit coupling of iron. The structure observed at 9.8° is the Fe Llη doublet (3s – $2p_{3/2,1/2}$ transitions). It is too less intense to be resolved.

Typically the WDS analysis of the Fe L range is made with a TlAP crystal, due to an adequate reticular distance or 2d-spacing of 1.28 nm and a relatively high reflectivity [11-13].

Thus, we compare in Figure 2 the Fe Lαβ doublet of the metal iron sample obtained with both kinds of multilayers and a TlAP crystal. The intensities are normalized to the same excitation conditions. As expected, the crystal gives the best spectral resolution and the lowest intensity while the multilayer gives the worst resolution but the highest intensity. The multilayer grating is intermediate, giving twice the intensity (measured from the peak height) of the crystal and improving the resolution by a factor less than two with respect to the multilayer. Thus the multilayer grating can be envisaged as a compromise between the multilayer and the crystal, in order to unambiguously resolve spectral features while collecting sufficiently large intensity.

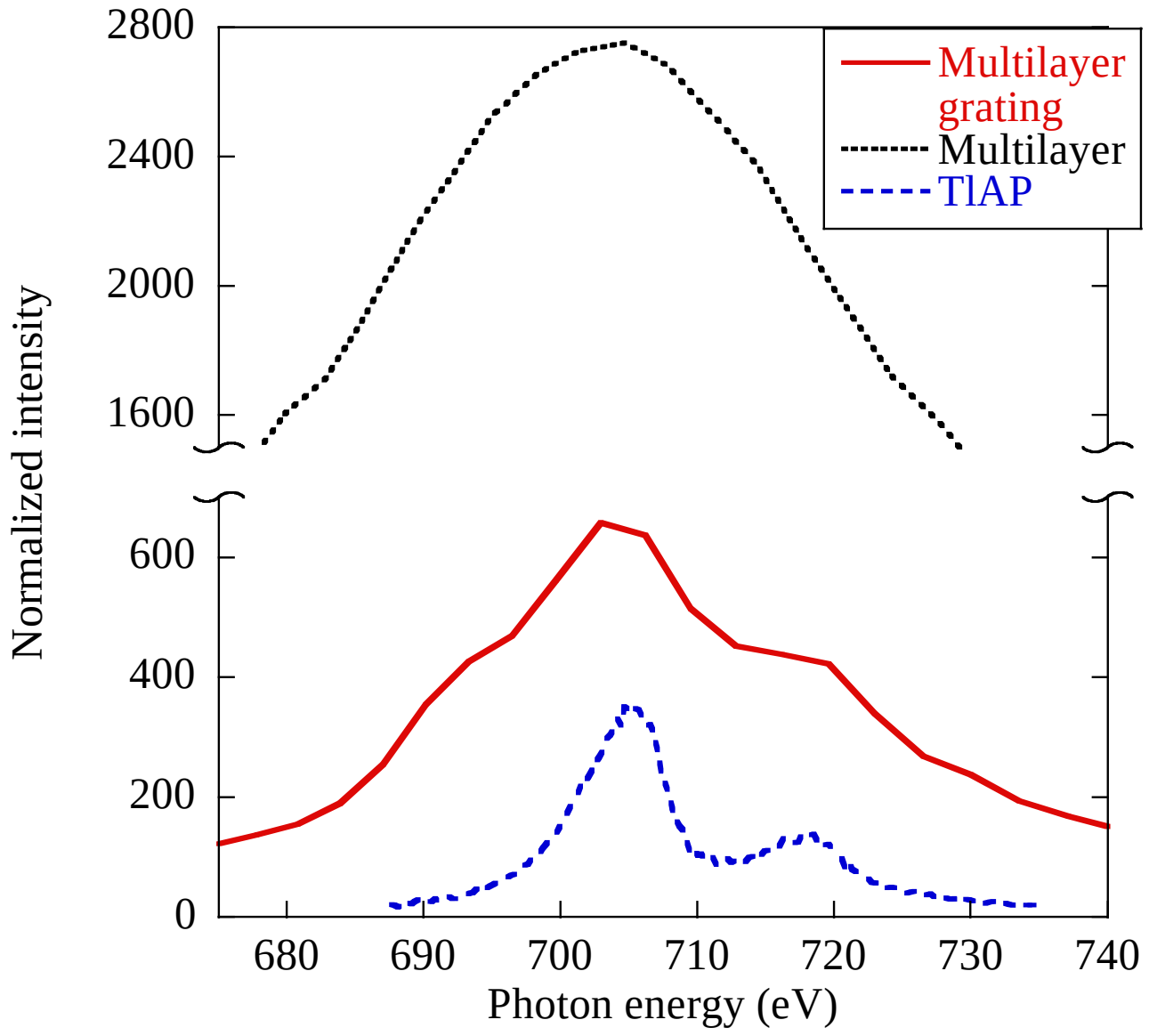

**Figure 2**: Comparison of the Fe Lαβ emission band of metal iron obtained with a $Mo/B_4C$ multilayer and the $Mo/B_4C$ multilayer grating and with a TlAP crystal.

The analysis of the mixed powder of LiF and iron ore with the multilayer grating is shown in Figure 3. A background is subtracted. The angular range is enlarged up to about 13° in order to observe the O Kα emission band (2p - 1s transition) coming from the oxygen atoms present in the iron ore. This emission is more than 0.5° wide. In the range around 9° two maxima are clearly observed: at lower grazing angles, the Fe Lαβ doublet; at higher angles, the F Kα emission (2p - 1s transition). Both emissions, separated by less than 0.3° (less than 30 eV), are easily resolved with the multilayer grating. Thus, in this case it would be rather easy to obtain the relative intensity of the Fe and F emissions by measuring either the intensity under each emission after decomposition of the spectrum into two components or simply the peak heights. Let us note that it is not possible to resolve Fe L from F K with the multilayer.

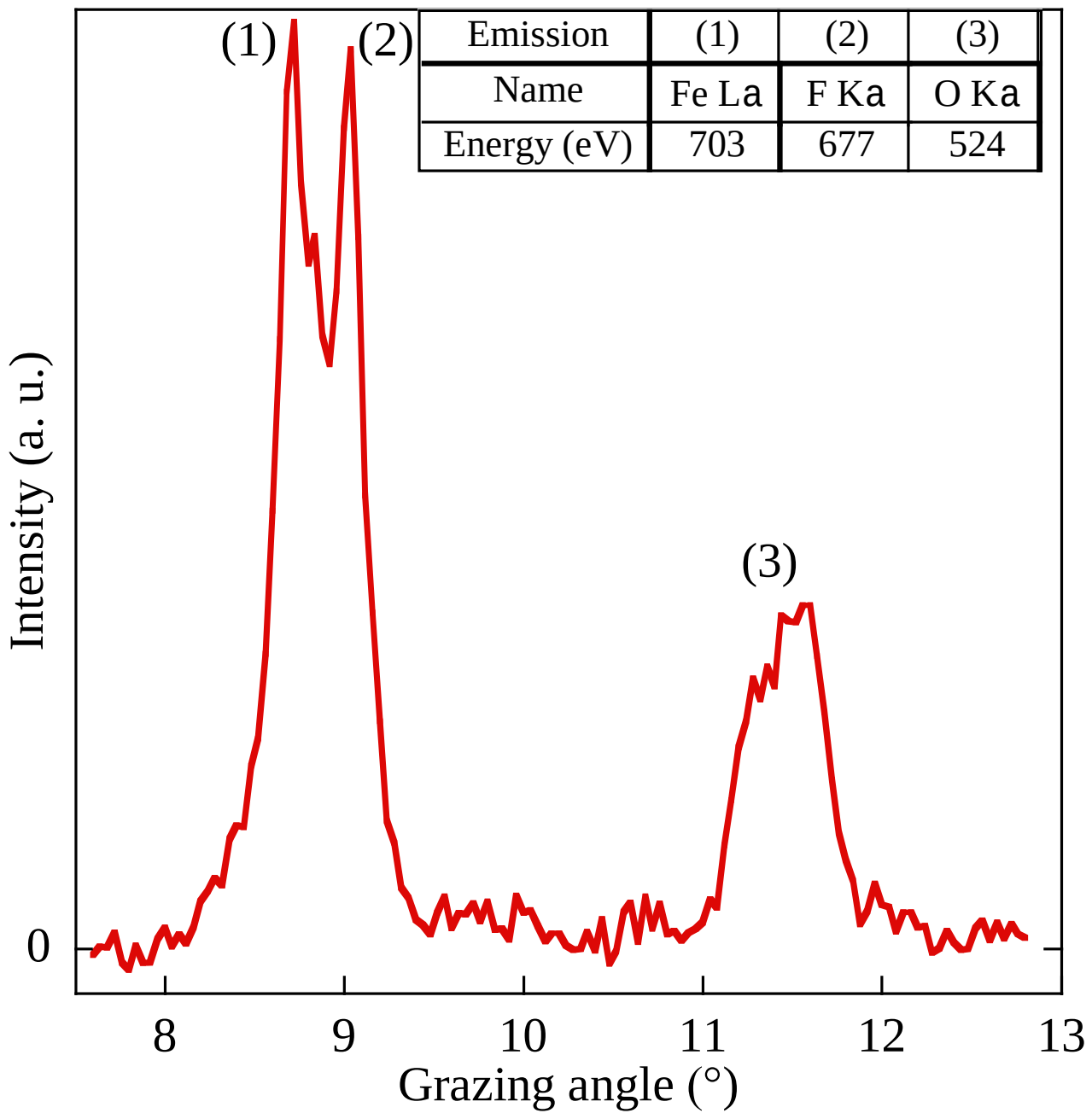

**Figure 3**: X-ray analysis of a mix of LiF and iron ore powders with the Mo/$B_4C$ multilayer grating.

## X-ray analysis in the C K range

We present in Figure 4 the C K$\alpha$ spectrum (2p - 1s transition) from the cellulose sample analysed with the multilayer and the multilayer grating. Both spectra are obtained in the same excitation conditions. When using the grating we observe:

- a small loss of reflectivity, not significant because the sample has evolved under the electron beam between the two analysis;
- a decrease of the FWMH by a factor of about 50%;
- a decrease of the background intensity by a factor 4;
- a shift of the maximum toward the high photon energies.

The FWMH obtained with the multilayer grating is 12.4 eV, still larger than the one that can be measured with grating spectrometer [14] around 7-8 eV. Considering that the instrumental broadening is negligible with a high resolution grating spectrometer, this means that in our case about half of the observed FWMH comes from the instrumental broadening introduced by the diffraction pattern of the multilayer grating. The decrease of the background intensity is due to the smaller total reflection angle when a multilayer grating is used [5]. The energy shift of the maximum is due to the fact that the conversion from the Bragg angles to the photon energies is made by using the Bragg law uncorrected from the refraction effect, **i.e. by taking into account the mutilayer period but not the refractive index that is different between the multilayer and the multilayer grating.**

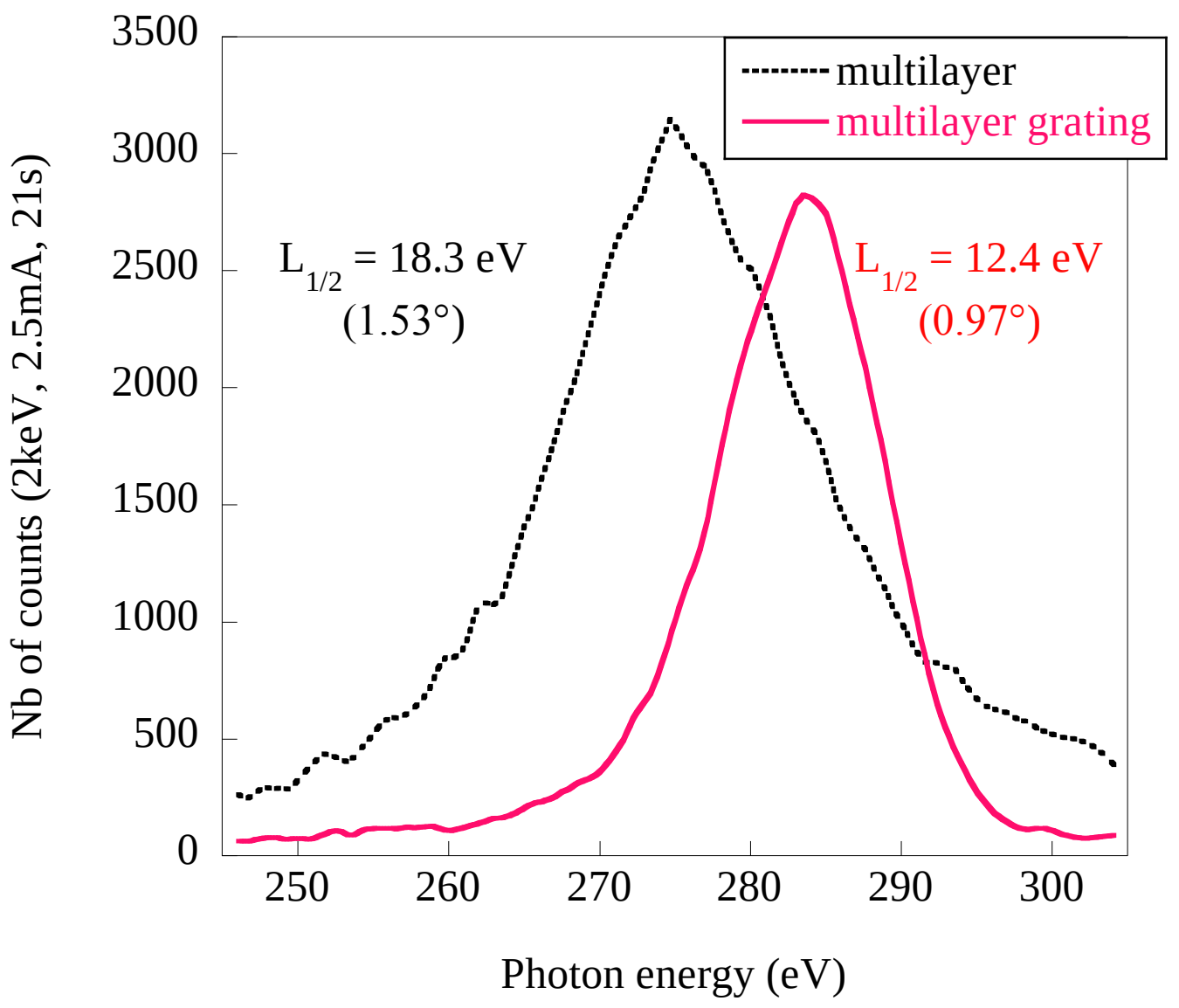

**Figure 4**: C Kα spectrum from the cellulose sample obtained with the multilayer and the multilayer grating.

We present in Figure 5 the C Kα spectrum of the cellulose sample compared to the one of the $B_4C$ thin film both measured with the multilayer grating. The spectra are normalized to their maximum and a linear background is subtracted. Since the C Kα emission is an emission band, its shape and width are sensitive to the chemical state of the carbon atoms within the sample. This is what is observed in figure 5, the FWMH of the carbide film (8.6 eV) is narrower than that of the cellulose (12.4 eV). In fact, due to its evolution under the electron beam, the cellulose sample should contain carbon atoms in many chemical states, leading to the broad emission band. It is also observed that the $B_4C$ spectrum presents an additional peak around 270 eV. For the analysis of this sample, the slit in front of the detector was opened too far and thus also accept a diffraction by the grating, leading to appearance of the parasitic peak.

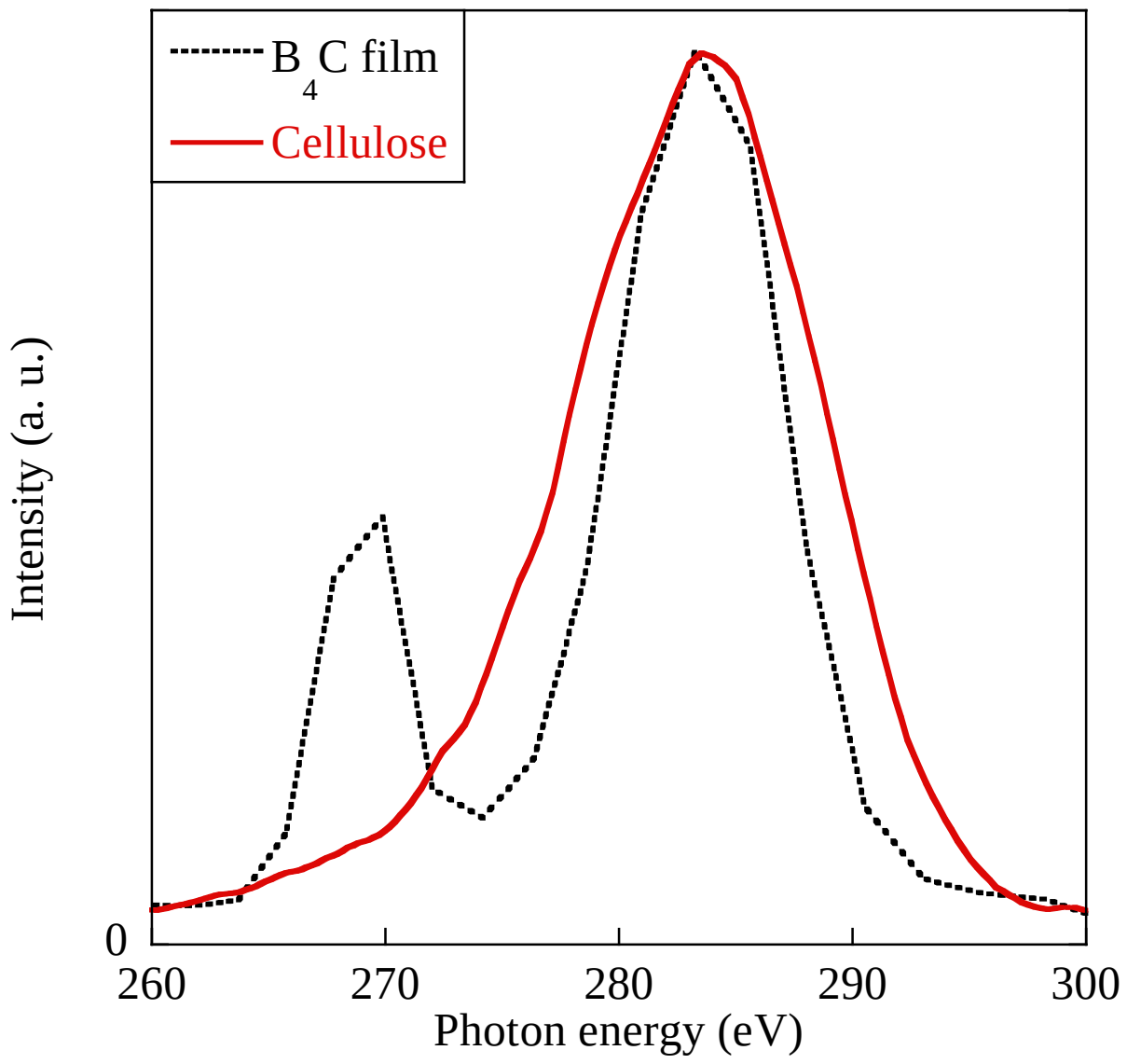

**Figure 5**: C Kα spectra of the cellulose and $B_4C$ samples obtained with the multilayer grating.

# Conclusion

Despite the fact that the diffracting systems, multilayer and multilayer grating, used in this study are not designed to work optimally in the C K and Fe L ranges, it has been possible to perform the x-ray analysis of various samples. We have demonstrated that the multilayer gives higher intensities, while the multilayer grating is superior in terms of spectral resolution. This enables to minimize line interference and so to obtain better intensity measurements when close lines are studied. As a first example, we have shown that it is possible to easily resolve the Fe L emission from the F Kα emission, and even to begin to resolve the two components of the Fe Lαβ doublet, separated by only 13 eV. As illustrated in the study of the C K emission band, the shape of the emission band is sensitive to the chemical state of the emitting atoms. The multilayer grating is also superior in terms of background and thus is useful to improve the detection limit. However, since the multilayer grating diffracts the x-rays using the multilayer structure, it is necessary to control the width of the slit (whose collected intensity and spectral resolution depend) in front of the detector in order to avoid parasitic emission coming from the grating structure.

# Acknowledgments

Dr. P. Hegeman from PANalytical is thanked for providing us with the cellulose and mixed powder samples.